\documentclass[twocolumn,showpacs,amsmath,amssymb,floatfix,aps,pra]{revtex4}
\usepackage{graphicx}

\def\sgn{{\rm sgn}}

\def\ep{\varepsilon}

\begin{document}

\title{Multiphoton antiresonance in large-spin systems}

\author{C. Hicke and M.~I.~Dykman} \affiliation{Department of Physics and Astronomy,
Michigan State University, East Lansing, MI 48824}

\date{\today}

\begin{abstract}
We study nonlinear response of a spin $S>1/2$ with easy-axis
anisotropy. The response displays sharp dips or peaks when the
modulation frequency is adiabatically swept through multiphoton
resonance. The effect is a consequence of a special symmetry of the
spin dynamics in a magnetic field for the anisotropy energy $\propto
S_z^2$. The occurrence of the dips or peaks is determined by the
spin state. Their shape strongly depends on the modulation
amplitude. Higher-order anisotropy breaks the symmetry, leading to
sharp steps in the response as function of frequency. The results
bear on the dynamics of molecular magnets in a static magnetic
field.

\end{abstract}
 \maketitle

\section{Introduction}
\label{sec:Intro}

Large-spin systems have been attracting much attention recently.
Examples are $S=3/2$ and $S=5/2$ Mn impurities in semiconductors and
Mn- and Fe-based molecular magnets with electron spin $S= 10$ and
higher. Nuclear spins $I=3/2$ have been also studied, and
radiation-induced quantum coherence between the spin levels was
observed \cite{Yusa2005}. An important feature of large-spin systems
is that their energy levels may be almost equidistant. A familiar
example is spins in a strong magnetic field in the case of a
relatively small magnetic anisotropy, where the interlevel distance
is determined primarily by the Larmor frequency. Another example is
low-lying levels of large-$S$ molecular magnets for small tunneling.
As a consequence of the structure of the energy spectrum, external
modulation can be close to resonance with many transitions at a
time. This should lead to coherent nonlinear resonant effects that
have no analog in two-level systems.

The effects of a strong resonant field on systems with nearly
equidistant energy levels have been studied for weakly nonlinear
oscillators. These studies concern both coherent effects, which
occur without dissipation
\cite{Larsen1976,Sazonov1976,Dmitriev1986a}, and incoherent effects,
in particular those related to the oscillator bistability and
transitions between coexisting stable states of forced vibrations.
In the absence of dissipation, a nonlinear oscillator may display
multiphoton antiresonance in which the susceptibility displays a dip
or a peak as a function of modulation frequency \cite{Dykman2005}.

In the present paper we study resonantly modulated spin systems with
$S>1/2$. Of primary interest are systems with uniaxial magnetic
anisotropy, with the leading term in the anisotropy energy of the
form of $-DS_z^2/2$. We show that the coherent response of such spin
systems displays peaks or dips when the modulation frequency
adiabatically passes through multiphoton resonances. The effect is
nonperturbative in the field amplitude. It is related to the special
conformal property of the spin dynamics in the semiclassical limit.
It should be noted that the occurrence of antiresonance for a spin
does not follow from the results for the oscillator. A spin can be
mapped onto a system of two oscillators rather than one; the
transition matrix elements for a spin and an oscillator are
different as are also the energy spectra.

We show that the coherent response of a spin is sensitive to terms
of higher order in $S_z$ in the anisotropy energy. In addition,
there is a close relation between the problem of resonant
high-frequency response of a spin and the problem of static spin
polarization transverse to the easy axis. Spin dynamics in a static
magnetic field has been extensively studied both theoretically and
experimentally
\cite{Chudnovsky1988,Garanin1991,Garg1993,Friedman1996,Friedman1997,Garanin1997a,Wernsdorfer1999,Garg1999,Villain2000,Wernsdorfer2006}.
One of the puzzling observations on magnetization switching in
molecular magnets, which remained unexplained except for the
low-order perturbation theory, is that the longitudinal magnetic
field at which the switching occurs is independent of the transverse
magnetic field \cite{Friedman1997}. The analysis presented below
provides an explanation which is nonperturbative
in the transverse field and also predicts the occurrence of peaks or
dips in the static polarization transverse to the easy axis as the
longitudinal magnetic field is swept through resonance.

\begin{figure}[h]
\includegraphics[width=3.2in]{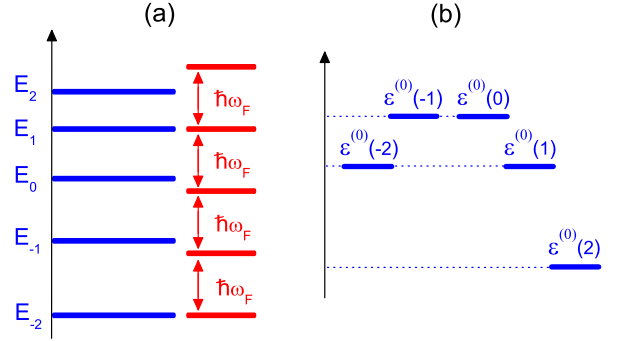}
\caption{(Color online). Three-photon resonance in a $S=2$ system in
the limit of a weak ac field. (a) Spin energy levels $E_m$ and
$n$-photon energies $n\hbar\omega_F$. (b) Quasienergies in the limit
of zero modulation amplitude, ${\ep^{(0)}( m)=E_ m- m\hbar
\omega_F}$; the pairwise degenerate levels correspond to one- and
three-photon resonance, respectively.} \label{fig:energies1}
\end{figure}

The onset of strong nonlinearity of the response due to near
equidistance of the energy levels can be inferred from
Fig.~\ref{fig:energies1}(a). It presents a sketch of the Zeeman
levels of a spin $E_{ m}$ ($-S\leq m\leq S$) in a strong magnetic
field along the easy magnetization axis $\hat{\bf z}$. The spin
Hamiltonian is
\begin{eqnarray}
\label{eq:hamiltonian_no_field}
 H_0 = \omega_0 S_z - \tfrac{1}{2}D S_z^2 \qquad (\hbar =1),
\end{eqnarray}
where $\omega_0$ is the Larmor frequency. For comparatively weak
anisotropy, $DS\ll \omega_0$, the interlevel distances $E_{ m+1}-E_{
m}$ are close to each other and change linearly with $ m$.

A transverse periodic field leads to transitions between neighboring
levels. An interesting situation occurs if the field frequency
$\omega_F$ is close to $\omega_0$ and there is multiphoton resonance
in the $m$th state: $N\omega_F$ coincides with the energy difference
$E_{ m+N}-E_{ m}$, $N>1$. The amplitude of the resonant $N$-photon
transition in this case is comparatively large, because the
transition goes via $N$ sequential one-photon virtual transitions
which are all almost resonant. Therefore one should expect a
comparatively strong multiphoton Rabi splitting already for a
moderately strong field.

A far less obvious effect occurs in the coherent response of the
system, that is in the magnetization at the modulation frequency or,
equivalently, the susceptibility. As we show, the expectation value
of the susceptibility displays sharp spikes at multiphoton
resonance. The shape of the spikes very strongly depends on the
field amplitude.

The paper is organized as follows. In Sec.~II we study the
quasienergy spectrum and the response of a spin with quadratic in
$S_z$ anisotropy energy. We show that, at multiphoton resonance, not
only multiple quasienergy levels are crossing pairwise, but the
susceptibilities in the resonating states are also crossing. In
Sec.~III we show that multiphoton transitions, along with level
repulsion, lead to the onset of spikes in the susceptibility and
find the shape and amplitude of the spikes as functions of frequency
and amplitude of the resonant field. In Sec.~IV we present a WKB
analysis of spin dynamics, which explains the simultaneous crossing
of quasienergy levels and the susceptibilities beyond perturbation
theory in the field amplitude. In Sec.~V the role of terms of higher
order in $S_z$ in the anisotropy energy is considered. Section ~VI
contains concluding remarks.

\section{Low-field susceptibility crossing}
\label{sec:susc_crossing}

\subsection{The quasienergy spectrum}
\label{subsec:quasienergy_spectrum}

We first consider a spin  with Hamiltonian $H_0$
(\ref{eq:hamiltonian_no_field}), which is additionally modulated by
an almost resonant ac field. The modulation can be described by
adding to $H_0$ the term $-S_xA\cos\omega_Ft$, where $A$
characterizes the amplitude of the ac field. As mentioned above, we
assume that the field frequency $\omega_F$ is close to $\omega_0$
and that $\omega_F,\omega_0\gg D, A,|\omega_F-\omega_0|$.

It is convenient to describe the modulated system in the
quasienergy, or Floquet representation. The Floquet eigenstates
$|\psi_{\ep}(t)\rangle$ have the property
$|\psi_{\ep}(t+\tau_F)\rangle = \exp(-i \ep
\tau_F)|\psi_{\ep}(t)\rangle$, where $\tau_F=2\pi/\omega_F$ is the
modulation period and $\ep$ is quasienergy. For resonant modulation,
quasienergy states can be found by changing to the rotating frame
using the canonical transformation $U(t)=\exp(-i\omega_FS_zt)$. In
the rotating wave approximation the transformed Hamiltonian is
\begin{eqnarray}
\label{eq:hamiltonian_rot_wave}
 &H = - \delta \omega S_z - \frac{1}{2}D S_z^2-\frac{1}{2}AS_x,\\
&\delta \omega = \omega_F - \omega_0.\nonumber
\end{eqnarray}
Here we disregarded fast-oscillating terms $\propto A\exp(\pm
2i\omega_Ft)$.

The Hamiltonian $H$ has a familiar form of the Hamiltonian of a spin
in a scaled static magnetic field with components $\delta\omega$ and
$A/2$ along the $\hat{\bf z}$ and $\hat{\bf x}$ axes, respectively.
Much theoretical work has been done on spin dynamics described by
this Hamiltonian in the context of molecular magnets.

The eigenvalues of $H$ give quasienergies of the modulated spin. In
the weak modulating field limit, $A\to 0$, the quasienergies are
shown in Fig.\ref{fig:energies1}(b). In this limit spin states are
the Zeeman states, i.e., the eigenstates $|m\rangle^{(0)}$ of $S_z$,
with $-S\leq m\leq S$. The interesting feature of the spectrum,
which is characteristic of the magnetic anisotropy of the form
$DS_z^2$, is that several states become simultaneously degenerate
pairwise for $A=0$ \cite{Friedman1997,Garanin1997a}. From
Eq.~(\ref{eq:hamiltonian_rot_wave}), the quasienergies $\ep^{(0)}(
m)$ and $\ep^{(0)}( m+N)$ are degenerate if the modulation frequency
is
\begin{equation}
\label{eq:pairwise_degeneracy}
 \delta\omega=\delta\omega_{m;N},\qquad \delta\omega_{m;N}= -D\left(m+\frac{1}{2}N\right).
 \end{equation}
The condition (\ref{eq:pairwise_degeneracy}) is simultaneously met
for all pairs of states with given $2 m+N$.  It coincides with the
condition of $N$-photon resonance $E_{ m+N}-E_{ m}=N\omega_F$. In
what follows $N$ can be positive and negative. There are $4S-1$
frequency values that satisfy the condition
(\ref{eq:pairwise_degeneracy}) for a given $S$.

The field $\propto A$ leads to transitions between the states
$|m\rangle^{(0)}$ and to quasienergy splitting. The level splitting
for the Hamiltonian (\ref{eq:hamiltonian_rot_wave}) was calculated
earlier \cite{Garanin1997a}. For multiphoton resonance, it is equal
to twice the multiphoton Rabi frequency $\Omega_R(m;N)$,
\begin{eqnarray}
\label{eq:Rabi_splitting}
 &&\Omega_R(m;N)=\left|A/2D\right|^{|N|}
 |D|\nonumber\\
 &&\times
 \left[\frac{(S+m+N)!(S-m)!}{(S+m)!(S-m-N)!}\right]^{\frac{1}{2}\sgn N}\frac{1}{2(|N|-1)!^2}
 \end{eqnarray}
The $N$-photon Rabi frequency (\ref{eq:Rabi_splitting}) is $\propto
A^{|N|}$, as expected. We note that the amplitude $A$ is scaled by
the anisotropy parameter $D$, which characterizes the
nonequidistance of the energy levels and is much smaller than the
Larmor frequency. Therefore $\Omega_R$ becomes comparatively large
already for moderately weak fields $A\sim D$.

We denote the true quasienergy states as $|\nu\rangle$, with integer
or half-integer $\nu$ such that $-S\leq \nu \leq S$. The
quasienergies $\ep_{\nu}$ do not cross. One can enumerate the states
$|\nu\rangle$ by thinking of them as the adiabatic states for slowly
increasing $\delta\omega$, starting from large negative
$\delta\omega$. For $-\delta\omega/DS \gg 1,|A|/D$ the states
$|\nu\rangle$ are very close to the Zeeman states $|
\nu\rangle^{(0)}$, with $\nu$ being the eigenvalue of $S_z$. This
then specifies the values of $\nu$ for all $\delta\omega$.

If the field is weak, the states $|\nu\rangle$ are close to the
corresponding Zeeman states, $|\nu\rangle\approx | m\rangle^{(0)}$,
for all $\delta\omega$ except for narrow vicinities of the resonant
values $\delta\omega_{m;N}$ given by
Eq.~(\ref{eq:pairwise_degeneracy}). The relation between the numbers
$\nu$ and $m$ for $|\nu\rangle\approx | m\rangle^{(0)}$ is
\begin{eqnarray}
\label{eq:nu_vs_m}
 \nu= m +
\sum^{\prime}\nolimits_N\theta\bigl(\delta\omega
-\delta\omega_{m;N}\bigr) \sgn N,
\end{eqnarray}
where $N$ runs from $-S- m$ to $S- m$; the term $N= 0$ is
eliminated, which is indicated by the prime over the sum;
$\theta(x)$ is the step function. In obtaining
Eq.~(\ref{eq:nu_vs_m}) we took into account that, for weak fields,
only neighboring quasienergy levels $\ep_{\nu}$ and $\ep_{\nu\pm 1}$
come close to each other. Eq.~\ref{eq:nu_vs_m}) defines the state
enumerating function $m(\nu)$.
\begin{figure}[h]
\includegraphics[width=2.9in]{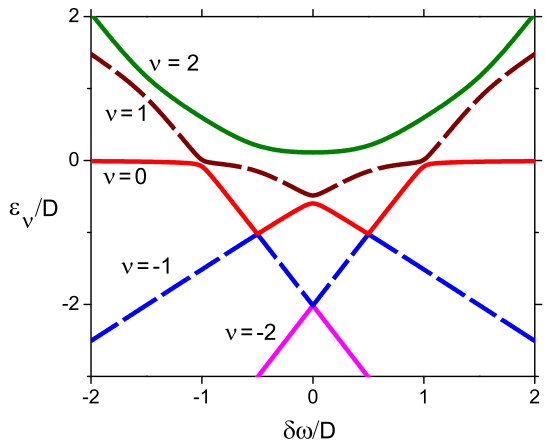}
\caption{(Color online). Quasienergy levels $\ep_{\nu}$ for a spin-2
system as functions of detuning $\delta \omega / D$ for the scaled
field amplitude $A/D=0.3$. The levels anticross pairwise at
multiphoton resonances given by Eq.~(\ref{eq:pairwise_degeneracy}).
The unperturbed quasienergies (the limit $A= 0$) correspond to
straight lines ${\ep^{(0)}( m)=-\delta\omega \, m - D m^2/2}$.}
 \label{fig:enumeration_scheme}
\end{figure}

The enumeration scheme and the avoided crossing of the quasienergy
levels are illustrated in Fig.~\ref{fig:enumeration_scheme}. For the
chosen $S=2$ the anticrossing occurs for 7 frequency values, as
follows from Eq.~(\ref{eq:pairwise_degeneracy}). The magnitude of
the splitting strongly depends on $N$: the largest splitting occurs
for one-photon transitions. It is also obvious from
Fig.~\ref{fig:enumeration_scheme} that several levels experience
anticrossing for the same modulation frequency.

\subsection{Susceptibility and quasienergy crossing}
\label{subsec:suscept_crossing}

Of central interest to us it the nonlinear susceptibility of the
spin. We define the dimensionless susceptibility $\chi_{\nu}$ in the
quasienergy state $|\nu\rangle$ as the ratio of the expectation
value of the appropriately scaled magnetization at the modulation
frequency to the modulation amplitude,
\begin{equation}
\label{eq:chi_definition}
 \chi_{\nu}(\omega_F)=\langle \nu|S_-|\nu\rangle/A.
 \end{equation}

In the weak field limit, $A\to 0$.
\begin{eqnarray}
\label{eq:weak_field}
 \chi_{\nu}(\omega_F)=\frac{ m(2\delta\omega +D  m) + D
 S(S+1)}{4(\delta\omega + D m)^2-D^2}
\end{eqnarray}
where $m$ and $ \nu$ are related by Eq.~(\ref{eq:nu_vs_m}); in fact,
Eq.~(\ref{eq:weak_field}) gives the susceptibility in the perturbed
to first order in $A$ Zeeman state $| m\rangle^{(0)}$.

A remarkable feature of Eq.~(\ref{eq:weak_field}) is the
susceptibility crossing at multiphoton resonance. The
susceptibilities in Zeeman states $|m\rangle^{(0)}$ and
$|m+N\rangle^{(0)}$ are equal where the unperturbed quasienergies of
these states are equal, $\ep^{(0)}(m)=\ep^{(0)}(m+N)$, i.e., where
the frequency detuning is $\delta\omega=\delta\omega_{m;N}$. In
terms of the adiabatic states $|\nu\rangle$, for such $\delta\omega$
we have from Eqs.~(\ref{eq:nu_vs_m}), (\ref{eq:weak_field})
$\chi_{\nu}(\omega_F)=\chi_{\nu'}(\omega_F)$ for $\nu'=\nu+\sgn N$.

A direct calculation shows that simultaneous crossing of the
susceptibilities and quasienergies occurs also in the fourth order
of the perturbation theory provided $N \geq 3$. Numerical
diagonalization of the Hamiltonian (\ref{eq:hamiltonian_rot_wave})
indicates that it persists in higher orders, too, until level
repulsion due to multiphoton Rabi oscillations comes into play.

The susceptibility $\chi_{\nu}$ is immediately related to the field
dependence of the quasienergy $\ep_{\nu}$. Since $\langle
\nu|S_+|\nu\rangle = \langle \nu|S_-|\nu\rangle$, from the explicit
form of the Hamiltonian (\ref{eq:hamiltonian_rot_wave}) we have
\begin{equation}
\label{eq:chi_vs_energy}
 \chi_{\nu}=-2A^{-1}\partial\ep_{\nu}/\partial A,
 \end{equation}
Simultaneous crossing of the susceptibilities and quasienergies
means that, for an $N$-photon resonance, the Stark shift of
resonating states is the same up to order $N-1$ in $A$; only in the
$N$th order the levels $\ep_{\nu}$ and $\ep_{\nu+\sgn N}$ become
split [by $2\Omega_R(m;N)$]. Respectively, the susceptibilities
$\chi_{\nu}$ and $\chi_{\nu +\sgn N}$ coincide up to terms $\propto
A^{|N|-3}$. The physical mechanism of this special behavior is
related to the conformal property of the spin dynamics, as explained
in Sec.~\ref{sec:classical}.

Equation (\ref{eq:weak_field}) does not apply in the case of
one-photon resonance, $N=1$: it gives $|\chi_{\nu}|\to \infty$ for
$\delta\omega\to \delta\omega_{m;1}$. This is similar to the case of
one-photon resonance in a two-level system, where the behavior of
the susceptibility is well understood beyond perturbation theory.
Interestingly, the lowest-order perturbation theory does not apply
also at exact two-photon resonance,
$\delta\omega=\delta\omega_{m;2}$, as discussed below, even though
Eq.~(\ref{eq:weak_field}) does not diverge.

\section{Antiresonance of the multiphoton response}
\label{sec:antiresonance}

\begin{figure}
\includegraphics[width=3.4in]{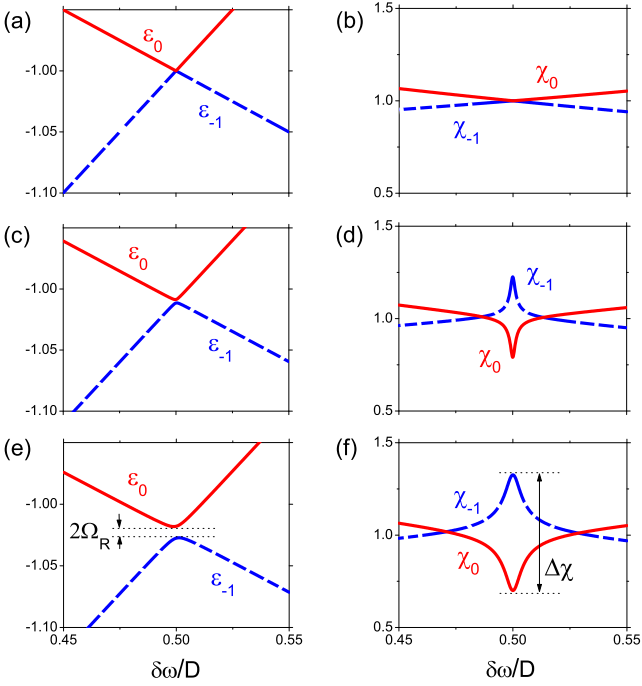}
\caption{(Color online). Level anticrossing and antiresonance of the
susceptibilities. The figure refers to a 3-photon resonance, $N=3$,
in an $S=2$ system. The involved quasienergy states are $\nu=-1$ and
$\nu=0$. The resonating Zeeman states for $A=0$ are $m=-2$ and $m=
1$ (the ground and 3rd excited state). Left and right panels show
the quasienergies $\ep_{\nu}$ and susceptibilities $\chi_{\nu}$ for
the same reduced field $A/D$. Panels (a) and (b), (c) and (d), and
(e) and (f) refer to $A/D=0, 0.2$, and $0.3$, respectively.}
 \label{fig:anticrossing}
\end{figure}

The field-induced anticrossing of quasienergy levels at multiphoton
resonance is accompanied by lifting the degeneracy of the
susceptibilities. It leads to the onset of a resonant peak and an
antiresonant dip in the susceptibilities as functions of frequency
$\delta\omega$. The behavior of the quasienergy levels and the
susceptibilities is seen from Fig.~\ref{fig:anticrossing}. For small
field amplitude $A$ the multiphoton Rabi frequency $\Omega_R\propto
A^{|N|}$ is small, the quasienergies of interest $\ep_{\nu}$ and
$\ep_{\nu+1}$ (with $m(\nu+1)-m(\nu)=N$) come very close to each
other at resonant $\delta\omega$, as do also the susceptibilities
$\chi_{\nu}$ and $\chi_{\nu+1}$.

With increasing $A$ the level splitting rapidly increases in a
standard way. The behavior of the susceptibilities is more
complicated. They cross, but sufficiently close to resonance they
repel each other, forming narrow dips (antiresonance) or peaks
(resonance). The widths and amplitudes of the dips/peaks display a
sharp dependence on the amplitude and frequency of the field.

For weak field it is straightforward to find the splitting of the
susceptibilities
\[\Delta\chi_{\nu;N}(\omega_F)=|\chi_{\nu}(\omega_F)-\chi_{\nu+\sgn N}(\omega_F)|\]
close to $N$-photon resonance between states $|m\rangle^{(0)}$ and
$|m+N\rangle^{(0)}$. In this region the frequency detuning from the
resonance
\begin{equation}
\label{eq:detuning_multiphoton}
 \Delta\omega(m;N)=N(\delta\omega-\delta\omega_{m;N})/2
\end{equation}
is small, $|\Delta\omega(m;N)|\lesssim \Omega_R(m;N)$. To the lowest
order in $A$ but for an arbitrary ratio
$\Omega_R(m;N)/|\Delta\omega(m;N)|$ the quasienergy states
$|\nu\rangle$ and $|\nu+\sgn N\rangle$ are linear combinations of
the states $|m\rangle^{(0)}$ and $|m+N\rangle^{(0)}$. Then from
Eq.~(\ref{eq:hamiltonian_rot_wave}) it follows that the splitting of
the quasienergies $\Delta\ep_{\nu;N}=|\ep_{\nu}-\ep_{\nu+\sgn N}|$
is
\begin{equation}
 \label{eq:splitting}
\Delta\ep_{\nu;N}=\left[\Delta\omega^2(m;N)+4\Omega_R^2(m;N)\right]^{1/2}.
\end{equation}

From this expression and Eqs.~(\ref{eq:Rabi_splitting}),
(\ref{eq:chi_vs_energy}) it follows that the susceptibility
splitting is
\begin{equation}
\label{eq:suscept_splitting}
 \Delta\chi_{\nu;N}=\frac{8|N|\Omega_R^2(m;N)}
 {A^2\left[\Delta\omega^2(m;N)+4\Omega_R^2(m;N)\right]^{1/2}}.
\end{equation}

The splitting $\Delta\chi_{\nu;N}$ as a function of frequency
$\delta\omega$ is maximal at $N$-photon resonance,
$\delta\omega=\delta\omega_{m;N}$. The half-width of the peak of
$\Delta\chi_{\nu;N}$ at half height is determined by the Rabi
splitting and is equal to $\sqrt{3}\Omega_R/N$. The peak is strongly
non-Lorentzian, it is sharper than the Lorentzian curve with the
same half-width. This sharpness is indeed seen in
Fig.~\ref{fig:anticrossing}. Our numerical results show that
Eq.~(\ref{eq:suscept_splitting}) well describes the splitting in the
whole frequency range $|\Delta\omega|\lesssim \Omega_R$.

\begin{figure}
\includegraphics[width=2.8in]{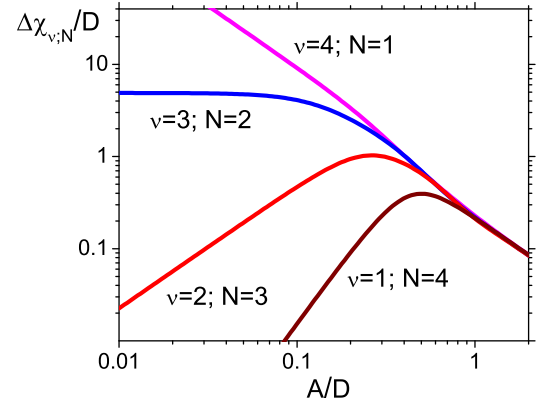}
\caption{(Color online). The multiphoton susceptibility splitting
for $S=2$. The curves refer to exact $N$-photon resonances, with
$N=1,\ldots, 4$, for transitions from the ground Zeeman state $m=-2$
to the excited states $m=-1,\ldots,2$, respectively.}
\label{fig:scaling}
\end{figure}

For small $A$, the susceptibility splitting is stronger than the
level repulsion. It follows from Eqs.~(\ref{eq:splitting}),
(\ref{eq:suscept_splitting}) that at exact $N$-photon resonance
$\Delta\ep \propto A^{|N|}$ whereas $\Delta\chi\propto A^{|N|-2}$.
This scaling is seen in Fig.~\ref{fig:scaling}. For $A/D\gg 1$, on
the other hand, the eigenstates $|\nu\rangle$ become close to the
eigenstates of a spin with Hamiltonian $-AS_x/2$. As a result, the
susceptibility splitting decreases with increasing $A$,
$|\Delta\chi_{\nu;N}|\propto A^{-1}$; the proportionality
coefficient here is independent of $N$. Therefore, for $N\geq 3$
$\Delta\chi_{\nu}$ displays a maximum as a function of $A$, as seen
from Fig.~\ref{fig:scaling}.

\subsection{Two-photon resonance}
\label{subsec:two_photon}

As mentioned above, the lowest order perturbation theory
(\ref{eq:weak_field}) does not describe resonant susceptibility for
two-photon resonance. Indeed, it follows from
Eq.~(\ref{eq:suscept_splitting}) that at exact resonance,
$\delta\omega=\delta\omega(m;2)$, the susceptibility splitting for
weak fields is
\begin{eqnarray}
\label{eq:chi2photon}
 \Delta\chi_{\nu;2}&=&D^{-1}\left[(S-m-1)(S-m)
 \right.\nonumber\\
 &&\times\left. (S+m+1)(S+m+2)\right]^{1/2}.
 \end{eqnarray}
This splitting is independent of $A$. The expression for the
susceptibility (\ref{eq:weak_field}) is also independent of $A$, yet
it does not lead to susceptibility splitting and therefore is
incorrect at two-photon resonance.

The inapplicability of the simple perturbation theory
(\ref{eq:weak_field}) is a consequence of quantum interference of
transitions, the effect known in the linear response of multilevel
systems \cite{DK_review84}. To the leading order in $A$, the
susceptibility is determined by the squared amplitudes of virtual
transitions to neighboring states. For a two-photon resonance,
$\delta\omega=\delta\omega_{m;2}$, the distances between the levels
involved in the transitions $|m\rangle^{(0)}\to |m+1\rangle^{(0)}$
and $|m+2\rangle^{(0)}\to |m+1\rangle^{(0)}$ are equal,
$\ep^{(0)}(m+1)-\ep^{(0)}(m)=\ep^{(0)}(m+1)-\ep^{(0)}(m+2)$.
Therefore the transitions resonate and interfere with each other.

To calculate the susceptibility it is necessary to start with a
superposition of states $|m\rangle^{(0)}$ and $|m+2\rangle^{(0)}$,
add the appropriately weighted amplitudes of transitions
$|m\rangle^{(0)}\to |m+1\rangle^{(0)}$ and $|m+2\rangle^{(0)}\to
|m+1\rangle^{(0)}$, and then square the result. This gives the
correct answer. The independence of the susceptibility splitting
from $A$ for two-photon resonance in the range of small $A$ as given
by Eq.~(\ref{eq:chi2photon}) is seen in Fig.~\ref{fig:scaling}.

\section{Susceptibility crossing for a semiclassical spin}
\label{sec:classical}

The analysis of the simultaneous level and susceptibility crossing
is particularly interesting and revealing for large spins and for
multiphoton transitions with large $N$. For $S\gg 1$ the spin
dynamics can be described in the WKB approximation. We will start
with the classical limit. In this limit it is convenient to use a
unit vector ${\bf s}={\bf S}/S$, with ${\bf s}\equiv
(s_x,s_y,s_z)\equiv (\sin\theta\cos\phi, \sin\theta\sin\phi,
\cos\theta)$, where $\theta$ and $\phi$ are the polar and azimuthal
angles of the vector ${\bf s}$. To the lowest order in $S^{-1}$
equations of motion for the spin components can be written as
\begin{eqnarray}
\label{eq:eqns_classical}
 \dot s_x =s_y(s_z+\mu),\qquad \dot s_y=-s_x(s_z+\mu)+fs_z,&&\\
 \dot s_z=-fs_y, \qquad f= A/2SD, \qquad \mu = \delta\omega/SD.&&\nonumber
 \end{eqnarray}
Here, overdot implies differentiation with respect to dimensionless
time $\tau = SDt$, that is, $\dot {\bf s}\equiv d{\bf s}/d\tau =
(SD)^{-1}d{\bf s}/dt$. Equations (\ref{eq:eqns_classical}) preserve
the length of the vector ${\bf s}$ and also the reduced Hamiltonian
$g=H/S^2D$,
\begin{eqnarray}
\label{eq:g_function}
 g\equiv g(\theta,\phi)=-\frac{1}{2}(s_z +\mu)^2 -fs_x.
 \end{eqnarray}
For convenience, we added to $g$ the term $-\mu^2/2$.

\begin{figure}[h]
\includegraphics[width=2.9in]{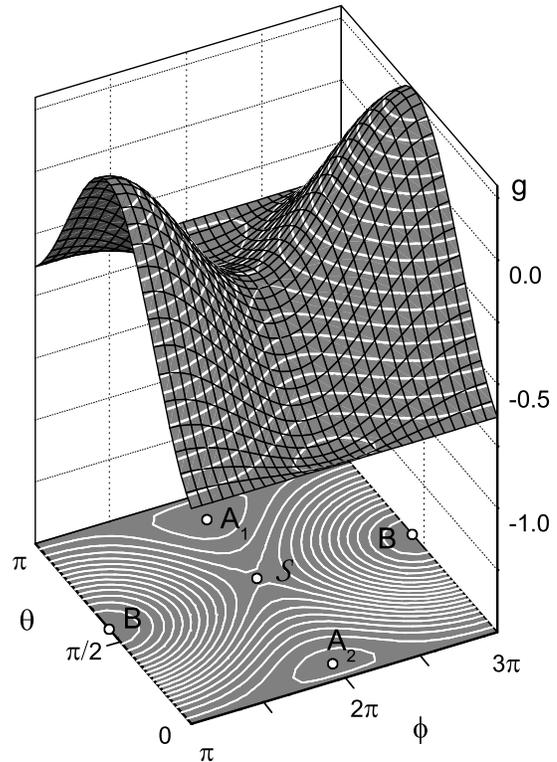}
\caption{The effective energy $g(\theta,\phi)$ as a function of the
polar and azimuthal angles of the classical spin $\theta$ and
$\phi$. The lines $g(\theta,\phi)=$~const describe classical spin
trajectories. The points $A_1$ and $A_2$ are the minima of $g$, $B$
is the maximum, and ${\cal S}$ is the saddle point. In the region
$g_{\cal S}>g > g_{A_1}$ there are two coexisting types of
trajectories. They lie on the opposite sides of the surface
$g(\theta,\phi)$ with respect to $g_{\cal S}$. The plot refers to
$\mu=0.125,\, f=0.3$.}\label{fig:trajectories}
\end{figure}

The effective energy $g(\theta,\phi)$ is shown in
Fig.~\ref{fig:trajectories}. Also shown in this figure are the
positions of the stationary states $\dot{\bf s}={\bf 0}$ and
examples of the phase trajectories described by
Eqs.~(\ref{eq:eqns_classical}).

An insight into the spin dynamics can be gained by noticing that $g$
has the form of the scaled free energy of an easy axis ferromagnet
\cite{LL_E&M}, with ${\bf s}$ playing the role of the magnetization
${\bf M}/M$, and with $\mu$ and $f$ being the reduced components of
the magnetic field along the easy axis $z$ and the transverse axis
$x$, respectively. In the region
\begin{equation}
\label{eq:hysteresis_LL}
 |f|^{2/3}+|\mu|^{2/3}< 1
 \end{equation}
the function $g$ has two minima, $A_1$ and $A_2$, a maximum $B$, and
a saddle point ${\cal S}$. We will assume that the minimum $A_2$ is
deeper than $A_1$, that is
\begin{equation}
\label{eq:shallow_min_condition}
 g_B > g_{\cal S} > g_{A_1} > g_{A_2}.
\end{equation}
As seen from Eqs.~(\ref{eq:eqns_classical}) and
(\ref{eq:g_function}) and Fig.~\ref{fig:trajectories}, for $f>0$ the
minima and the saddle point are located at $\phi=0$ and the maximum
is at $\phi=\pi$; the case $f<0$ corresponds to a replacement
$\phi\to \phi+\pi$. On the boundary of the hysteresis region
(\ref{eq:hysteresis_LL}) the shallower minimum $A_1$ merges with the
saddle point ${\cal S}$.

In the case of an easy-axis ferromagnet with free energy $g$, the
minima of $g$ correspond to coexisting states of magnetization
within the hysteresis region (\ref{eq:hysteresis_LL}). For
multiphoton absorption $g$ is the scaled quasienergy, not free
energy, and stability is determined dynamically by balance between
relaxation and high-frequency excitation. One can show that, for
relevant energy relaxation mechanisms, the system can have
coexisting stable stationary states inside and outside the region
(\ref{eq:hysteresis_LL}). The states correspond to one or both
minima and/or the maximum of $g$; for small damping the actual
stable states are slightly shifted away from the extrema of $g$ on
the ($\theta,\phi$)-plane. We will not discuss relaxation effects in
this paper.

\subsection{Conformal property of classical trajectories}
\label{subsec:symmetry_trajectories}

Dynamical trajectories of a classical spin on the plane
$(\theta,\phi)$ are the lines $g(\theta,\phi)=$~const. They are
either closed orbits around one of the minima $A_1, A_2$ or the
maximum $B$ of $g$, or open orbits along the $\phi$ axis, see
Fig.~\ref{fig:trajectories}. On the Bloch sphere ${\bf s}^2=1$,
closed orbits correspond to precession of the unit vector ${\bf s}$
around the points ${\bf s}_{A_1}, {\bf s}_{A_2}$, or ${\bf s}_B$, in
which ${\bf s}$ does not make a complete turn around the polar axis.
Open orbits correspond to spinning of ${\bf s}$ around the polar
axis accompanied by oscillations of the polar angle $\theta$. Even
though the spin has 3 components, the spin dynamics is the dynamics
with one degree of freedom, the orbits on the Bloch sphere do not
cross.

An important feature of the dynamics of a classical spin in the
hysteresis region is that, for each $g$ in the interval $(g_{A_1},
g_{\cal S})$, the spin has two coexisting orbits, see
Fig.~\ref{fig:trajectories}. One of them corresponds to spin
precession around ${\bf s}_{A_1}$. It can be a closed loop or an
open trajectory around the point $A_1$ on the $(\theta,\phi)$-plane.
The other is an open trajectory on the opposite side of the
$g$-surface with respect to the saddle point. We will classify them
as orbits of type I and II, respectively.

We show in Appendix that classical equations of motion can be solved
in an explicit form, and the time dependence ${\bf s}(\tau)$ is
described by the Jacobi elliptic functions. The solution has special
symmetry. It is related to the conformal property of the mapping of
$s_z$ onto $\tau$. The major results of the analysis are the
following features of the trajectories ${\bf s}(\tau)$ of types I
and II: for equal $g$, (i) their dimensionless oscillation
frequencies $\omega(g)$ are equal to each other, and (ii) the period
averaged values of the component $s_x(\tau)$ are equal, too,
\begin{equation}
\label{eq:class_traj_summary}
 \omega_I(g)= \omega_{II}(g),\qquad \langle s_x(\tau)\rangle_I= \langle
 s_x(\tau)\rangle_{II}.
\end{equation}
Here, the subscripts I and II indicate the trajectory type. The
angular brackets $\langle\ldots\rangle$ imply period averaging on a
trajectory with a given $g$.

The quantity $\langle s_x(\tau)\rangle$ gives the classical response
of the spin to the field $\propto A$.
Equation~(\ref{eq:class_traj_summary}) shows that this response is
equal for the trajectories with equal values of the effective
Hamiltonian function $g$. This result holds for any field amplitude
$A$, it is by no means limited to small $A/D$ where the perturbation
theory in $A$ applies.

\subsection{The WKB picture in the neglect of tunneling}
\label{subsec:WKB_picture}

In the WKB approximation, the values of quasienergy $\ep_{\nu}$ in
the neglect of tunneling can be found by quantizing classical orbits
$g(\theta,\phi)=$~const, see Ref.~\onlinecite{Garg2004} and papers
cited therein. Such quantization should be done both for orbits of
type I and type II, and we classify the resulting states as the
states of type I and II, respectively. The distance between the
states of the same type in energy units is $\hbar\omega(g)SD$
\cite{LL_QM81}. Transitions between states of types I and II with
the same $g$ are due to tunneling.

If we disregard tunneling, the quasienergy levels of states I and II
will cross, for certain values of $\mu$. Remarkably, if two levels
cross for a given $\mu$, then all levels in the range $g_{A_1}< g <
g_{\cal S}$ cross pairwise. This is due to the fact that the
frequencies $\omega(g)$ and thus the interlevel distances for the
two sets of states are the same, see
Eq.~(\ref{eq:class_traj_summary}). Such simultaneous degeneracy of
multiple pairs of levels agrees with the result of the low-order
quantum perturbation theory in $A$ and with numerical calculations.

In the WKB approximation, the expectation value of an operator in a
quantum state is equal to the period-averaged value of the
corresponding classical quantity along the appropriate classical
orbit \cite{LL_QM81}. Therefore if semiclassical states of type I
and II have the same $g$, the expectation values of the operator
$S_x$ in these states are the same according to
Eq.~(\ref{eq:class_traj_summary}). Thus, the WKB theory predicts
that, in the neglect of tunneling, there occurs simultaneous
crossing of quasienergy levels and susceptibilities for all pairs of
states with quasienergies between $g_{A_1}$ and $g_{\cal S}$. This
is in agreement with the result of the perturbation theory in $A$
and with numerical calculations. However, we emphasize that the WKB
theory is not limited to small $A$, and the WKB analysis reveals the
symmetry leading to the simultaneous crossing of quasienergy levels
and the susceptibilities.

Tunneling between semiclassical states with equal $g$ leads to level
repulsion and susceptibility antiresonance. The level splitting
$2\Omega_R$ can be calculated by appropriately generalizing the
standard WKB technique, for example as it was done in the analysis
of tunneling between quasienergy states of a modulated oscillator
\cite{Dmitriev1986a}. Then the resonant susceptibility splitting can
be found from Eq.~(\ref{eq:chi_vs_energy}). The corresponding
calculation is beyond the scope of this paper.

\section{Degeneracy lifting by higher order terms in $S_z$}
\label{sec:s_z_4}

The simultaneous crossing of quasienergy levels and susceptibilities
in the neglect of tunneling is a feature of the spin dynamics
described by Hamiltonian (\ref{eq:hamiltonian_rot_wave}).
Higher-order terms in $S_z$ lift both this degeneracy and the
property that many quasienergy levels are pairwise degenerate for
the same values of the frequency detuning $\delta\omega$. The effect
is seen already if we incorporate the term $S_z^4$ in the anisotropy
energy, i.e. for a spin with Hamiltonian
\begin{equation}
\label{eq:s_z_4}
 \tilde H = H-\frac{1}{4}GS_z^4.
\end{equation}
The Hamiltonian $\tilde H$ is written in the rotating wave
approximation, $H$ is given by Eq.~(\ref{eq:hamiltonian_rot_wave}),
and $G$ is the parameter of quartic anisotropy. The terms $S_x^2,
S_y^2$ in the spin anisotropy energy do not show up in $\tilde H$
even if they are present in the spin Hamiltonian $H_0$ but the
corresponding anisotropy parameters are small compared to
$\omega_0$. In the rotating frame these terms renormalize the
coefficient at $S_z^2$ and lead to fast oscillating terms $\propto
S_{\pm}^2\exp(\pm 2i\omega_Ft)$ that we disregard.

Multiple pairwise degeneracy occurs where the condition on Zeeman
quasienergies $\ep^{(0)}(m)= \ep^{(0)}(m')$ is simultaneously met
for several pairs $(m,m')$. For $G\neq 0$ this happens only for
$\delta\omega=0$, that is when the modulation frequency $\omega_F$
is equal to the Larmor frequency $\omega_0$. In this case the
resonating Zeeman states are $|m\rangle^{(0)}$ and
$|-m\rangle^{(0)}$ with the same $m$. The susceptibilities of these
states are equal by symmetry with respect to reflection in the plane
$(x,y)$.

\begin{figure}
\includegraphics[width=3.4in]{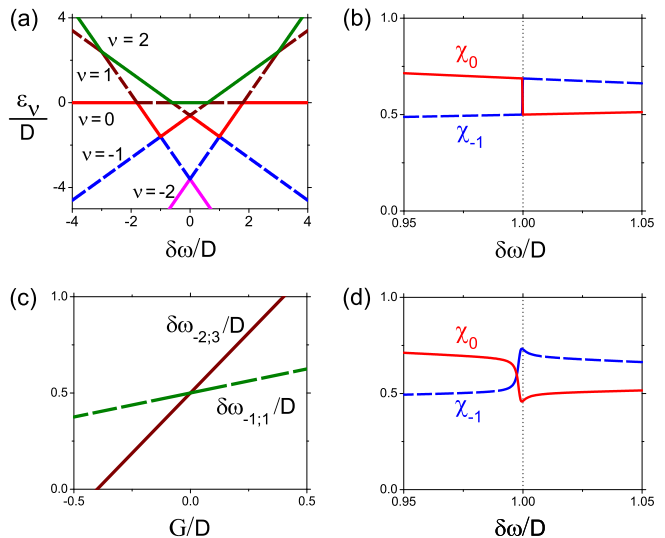} \caption{(Color online).
Quasienergy and susceptibility switching for a spin $S=2$ with
quartic in $S_z$ anisotropy. Panels (a), (b), and (d) refer to the
dimensionless quartic anisotropy parameter $G/D=0.4$ in
Eq.~(\ref{eq:s_z_4}). Panels (b) and (d) refer to the 3-photon
resonance $|-2\rangle^{(0)}\to |1\rangle^{(0)}$ with the scaled
modulation amplitude $A/D\to 0$ and $A/D=0.3$, respectively; the
dotted line shows the position of the resonance $\delta\omega/D=1$.
Panel (c) shows the dependence of the resonant frequency detuning
$\delta\omega_{m;N}$ on the higher-order anisotropy parameter $G$ in
the limit $A\to 0$.} \label{fig:quartic}
\end{figure}

$N$-photon resonance for nonzero $G$ and $\omega_F\neq \omega_0$
occurs generally only for one pair of states $|m\rangle^{(0)}$ and
$|m+N\rangle^{(0)}$. This is seen from panel (a) in
Fig.~\ref{fig:quartic}. With increasing $|G|$ the difference in the
resonant values of frequency increases, as seen from panel (c) in
the same figure.

The susceptibilities in resonating states are different in the
weak-field limit. When the frequency $\omega_F$ adiabatically goes
through resonance, there occurs an interchange of states, for weak
field $A$: if the state $|\nu\rangle$ was close to $|m\rangle^{(0)}$
on one side of resonance, it becomes close to $|m+N\rangle^{(0)}$ on
the other side. Respectively, the susceptibility $\chi_{\nu}$
sharply switches from its value in the state $|m\rangle^{(0)}$ to
its value in the state $|m+N\rangle^{(0)}$.

Susceptibility switching is seen in panels (b) and (d) in
Fig.~\ref{fig:quartic}. For a weak field the frequency range where
the switching occurs is narrow and the switching is sharp (vertical,
in the limit $A\to 0$). As the modulation amplitude $A$ increases
the range of frequency detuning $\delta\omega$ over which the
switching occurs broadens. In addition, for small $G$ the
susceptibility displays spikes. They have the same nature as for
$G=0$. However, they are much less pronounced, as seen from the
comparison of panel (d) in Fig.~\ref{fig:quartic} and panel (f) in
Fig.~\ref{fig:anticrossing} which refer to the same value of $A/D$.

\section{Conclusions}
\label{sec:conclusions}

In this paper we have considered a large spin with an easy axis
anisotropy. The spin is in a strong magnetic field along the easy
axis and is additionally modulated by a transverse field with
frequency $\omega_F$ close to the Larmor frequency $\omega_0$. We
have studied the coherent resonant response of the spin. It is
determined by the expectation value of the spin component transverse
to the easy axis. We are interested in multiphoton resonance where
$N\omega_F$ coincides or is very close to the difference of the
Zeeman energies $E_{m+N}^{(0)}-E_m^{(0)}$ in the absence of
modulation.

The major results refer to the case where the anisotropy energy is
of the form $-DS_z^2/2$. In this case not only the quasienergies of
the resonating Zeeman states $|m\rangle^{(0)}$ and
$|m+N\rangle^{(0)}$ cross at multiphoton resonance, but the
susceptibilities in these states also cross, in the weak-modulation
limit. Such crossing occurs simultaneously for several pairs of
Zeeman states. As the modulation amplitude $A$ increases, the levels
are Stark-shifted and the susceptibilities are also changed.
However, as long as the Rabi splitting due to resonant multiphoton
transitions (tunneling) can be disregarded, for resonant frequency
the quasienergy levels remain pairwise degenerate and the
susceptibilities remain crossing. We show that this effect is
nonperturbative in $A$, it is due to the special conformal property
of the classical spin dynamics.

Resonant multiphoton transitions lift the degeneracy of quasienergy
levels, leading to a standard level anticrossing. In contrast, the
susceptibilities as functions of frequency cross each other.
However, near resonance they display spikes. The spikes of the
involved susceptibilities point in the opposite direction, leading
to decrease (antiresonance) or increase (resonance) of the response.
They have a profoundly non-Lorentzian shape
(\ref{eq:suscept_splitting}), with width and height that strongly
depend on $A$. The spikes can be observed by adiabatically sweeping
the modulation frequency through a multiphoton resonance. If the
spin is initially in the ground state, a sequence of such sweeps
allows one to study the susceptibility in any excited state provided
the relaxation time is long enough.

The behavior of the susceptibilities changes if terms of higher
order in $S_z$ in the anisotropy energy are substantial. In this
case crossing of quasienergy levels is not accompanied by crossing
of the susceptibilities in the limit $A\to 0$. Resonant multiphoton
transitions lead to step-like switching between the branches of the
susceptibilities of the resonating Zeeman states. Still, the
susceptibilities display spikes as functions of frequency for a
sufficiently strong modulating field.

The results of the paper can be applied also to molecular magnets in
a static magnetic field. The spin Hamiltonian in the rotating wave
approximation (\ref{eq:hamiltonian_rot_wave}) is similar to the
Hamiltonian of a spin in a comparatively weak static field, with the
Larmor frequency $\delta\omega$ of the same order as the anisotropy
parameter $D$. The susceptibility then characterizes the response to
the field component transverse to the easy axis. Quasienergies
$\ep^{(0)}(m)$ are now spin energies in the absence of the
transverse field, and instead of multiphoton resonance we have
resonant tunneling. Our results show that a transverse field does
not change the value of the longitudinal field for which the energy
levels cross, in the neglect of tunneling. This explains the
experiment \cite{Friedman1997} where such behavior was observed.

In conclusion, we have studied multiphoton resonance in large-spin
systems. We have shown that the coherent nonlinear response of the
spin displays spikes when the modulation frequency goes through
resonance. The spikes have non-Lorentzian shape which strongly
depends on the modulation amplitude. The results bear on the
dynamics of molecular magnets in a static magnetic field and provide
an explanation of the experiment.

We acknowledge insightful discussions with B.L.~Altshuler and
A.~Kamenev. This work was supported by the NSF through grants No.
ITR-0085922 and PHY-0555346.

\appendix
\section{Symmetry of classical spin dynamics: a feature of the conformal mapping}
\label{sec:appendix}

Classical equations of motion for the spin components
(\ref{eq:eqns_classical}) can be solved in the explicit form, taking
into account that ${\bf s}^2=1$ and that $g(\theta,\phi)=$~const on
a classical trajectory. For time evolution of the $z$-component of
the spin we obtain
\begin{eqnarray}
\label{eq:s_z_explicit}
 s_z(\tau)=\frac{r_2(r_1-r_3)-r_3(r_1-r_2){\rm sn}^2(u;m_J)}
 {r_1-r_3-(r_1-r_2){\rm sn}^2(u;m_J)}
\end{eqnarray}
where $r_{1} > r_2 >r_3 >r_4$ are the roots of the equation
\begin{equation}
\label{eq:roots}
 \left[(r+\mu)^2+2g\right]^2+4f^2(r^2-1)=0
\end{equation}
and ${\rm sn}(u;m_J)$ is the Jacobi elliptic function. The argument
$u$ and the parameter $m_J$ are
\begin{eqnarray}
\label{eq:Jacobi_argument}
 u&=&\tilde\omega\tau,\qquad \tilde\omega=\frac{1}{4}\left[(r_1-r_3)(r_2-r_4)\right]^{1/2},
 \nonumber\\
m_J&=&(r_1-r_2)(r_3-r_4)/(r_1-r_3)(r_2-r_4),
\end{eqnarray}
Equation (\ref{eq:s_z_explicit}) describes an orbit which, for a
given $g$, oscillates between $s_z=r_1$ and $s_z=r_2$; the
corresponding oscillations of $s_x,s_y$ can be easily found from
Eqs.~(\ref{eq:eqns_classical}), (\ref{eq:g_function}).

Oscillations of $s_z$ between $r_3$ and $r_4$ for the same $g$ are
also described by Eq.~(\ref{eq:s_z_explicit}) provided one replaces
$u \to u+K(m_J) + iK'(m_J)$, where $K(m_J)$ is the elliptic integral
and $K'(m_J)=K(1-m_J)$. Clearly, both types of oscillations have the
same period over $\tau$ equal to $2K(m_J)/\tilde\omega$. They
correspond, respectively, to the trajectories of types II and I in
Fig.~\ref{fig:trajectories} that lie on different sides of
$g(\theta,\phi)$-surface. As a consequence, the vibration
frequencies for the corresponding trajectories $\omega_{II}(g)$ and
$\omega_{I}(g)$ are the same. This proves the first relation in
Eq.~(\ref{eq:class_traj_summary}).

The Jacobi elliptic functions are double periodic, and therefore
$s_z$ is also double periodic,
\begin{equation}
\label{eq:s_z_periodicity}
 s_z(\tau)=s_z\left[\tau+\tilde\omega^{-1}(2nK+2imK')\right]
\end{equation}
with integer $n,m$. Ultimately, this is related to the fact that
equations of motion (\ref{eq:eqns_classical}) after simple
transformations can be put into a form of a Schwartz-Christoffel
integral that performs conformal mapping of the half-plane
Im~$s_z>0$ onto a rectangle on the $u$-plane.
\begin{figure}[ht]
\centering
\includegraphics[width=2.8in]{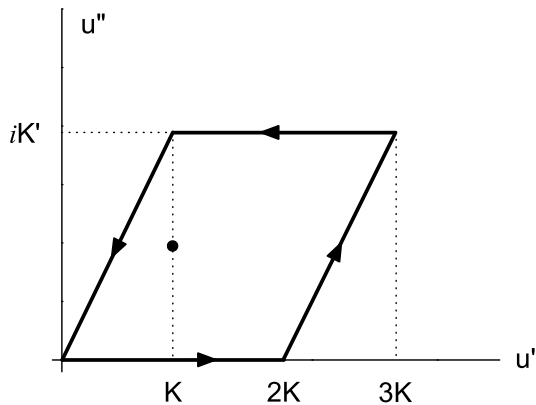}
\caption{The contour of integration in the $u\propto\tau$ plane. The
horizontal parts correspond to two trajectories ${\bf s}(\tau)$ with
the same $g$. The values of ${\bf s}(\tau)$ on the tilted parts of
the parallelogram are the same. The plot refers to $\mu=0.125,
g=-0.366$.} \label{fig:contour}
\end{figure}
We will show now that the mapping has a special property that leads
to equal period-averaged values of $s_x(\tau)$ on trajectories of
different types but with the same $g$. Because $s_z(\tau)$ is double
periodic, cf. Eq.~(\ref{eq:s_z_periodicity}), so is also the
function $s_x(\tau) = -(2f)^{-1}\left[2g+(s_z(\tau)+\mu)^2\right]$.
Keeping in mind that the transformation $u \to u+K(m_J) + iK'(m_J)$
moves us from a trajectory with a given $g$ of type I to a
trajectory of type II, we can write the difference of the
period-averaged values of $s_x(\tau)$ on the two trajectories as
\begin{eqnarray}
\label{eq:period_averaging}
 \langle s_x(\tau)\rangle_I - \langle s_x(\tau)\rangle_{II} =
 \frac{\omega(g)}{2\pi\tilde\omega}\oint\nolimits_C s_x\,du
\end{eqnarray}
where the contour $C$ is a parallelogram on the $u$-plane with
vortices at $0,2K, 3K+iK', K+iK'$. It is shown in
Fig.~\ref{fig:contour}.

An important property of the mapping (\ref{eq:s_z_explicit}) is that
$s_z(\tau)$ has one simple pole inside the contour $C$, as marked in
Fig.~\ref{fig:contour}. Respectively, $s_x(\tau)$ has a second-order
pole. The explicit expression (\ref{eq:s_z_explicit}) allows one to
find the corresponding residue. A somewhat cumbersome calculation
shows that it is equal to zero. This shows that the period-averaged
values of $s_x$ on the trajectories with the same $g$ coincide, thus
proving the second relation in Eq.~(\ref{eq:class_traj_summary}).


\end{document}